\documentclass[
reprint,
superscriptaddress,
showpacs,
 amsmath,amssymb,
 aps,
pra,
]{revtex4-1}

\usepackage{graphicx}% Include figure files
\usepackage{dcolumn}% Align table columns on decimal point
\usepackage{bm}% bold math
\usepackage{color}
\begin{document}

\preprint{}

\date{}
 
\title{Optical Properties of Vanadium in 4H Silicon Carbide for Quantum Technology}
 %Title of paper
\author{L. Spindlberger}
\affiliation{Institute for Semiconductor and Solid-State Physics, Johannes Kepler University Linz, Altenbergerstraße 69, 4040 Linz, Austria}
\author{A. Cs\'or\'e}
\author{G. Thiering}
\affiliation{Institute for Solid State Physics and Optics, Wigner Research Center for Physics, Hungarian Academy of Sciences, P.O. Box 49, H-1525 Budapest, Hungary}
\author{S. Putz}
\affiliation{Vienna Center for Quantum Science and Technology, Universit\"at Wien, Boltzmanngasse 5, 1090 Vienna}
\author{R. Karhu} 
\author{J. Ul Hassan}
\author{N. T. Son}
\affiliation{Department of Physics, Chemistry and Biology, Link{\"o}ping University, SE-58183 Link{\"o}ping, Sweden}
\author{T. Fromherz}
\affiliation{Institute for Semiconductor and Solid-State Physics, Johannes Kepler University Linz, Altenbergerstraße 69, 4040 Linz, Austria}
\author{A. Gali}
\affiliation{Institute for Solid State Physics and Optics, Wigner Research Center for Physics, Hungarian Academy of Sciences, P.O. Box 49, H-1525 Budapest, Hungary}
\affiliation{Department of Atomic Physics, Budapest, University of Technology and Economics, Budafoki {\'u}t 8, H-1111, Budapest, Hungary}
\author{M. Trupke}
\email{All correspondence should be addressed to michael.trupke@univie.ac.at}
\affiliation{Vienna Center for Quantum Science and Technology, Universit\"at Wien, Boltzmanngasse 5, 1090 Vienna}

\date{\today}

\begin{abstract}
We study the optical properties of tetravalent vanadium impurities in 4H silicon carbide (4H SiC). Emission from two crystalline sites is observed at wavelengths of 1.28 $\mu$m and 1.33 $\mu$m, with optical lifetimes of 163 ns and 43 ns. Group theory and {\it ab initio} density functional supercell calculations enable unequivocal site assignment and shed light on the spectral features of the defects. We conclude with a brief outlook on applications in quantum photonics. 
 \end{abstract}
%\pacs{76.30.Mi, 07.55.Ge, 03.67.Pp}% PACS, the Physics and Astronomy
                             % Classification Scheme.

\maketitle

%TC:endignore 
\section{Introduction}
Defects in semiconductors such as silicon, diamond and silicon carbide have advanced to the forefront of candidates for the implementation of quantum bits and sensors, as they present attractive features such as long quantum coherence lifetimes and strong optical transitions  \cite{Doherty2013,Rondin2014,Kucsko2013,Dolde2011,Neumann2013,
Rogers2014,Becker2016,Sipahigil2016,Christle2015,Widmann2015,Fuchs2015,Bosma2018,Atatuere2018}. In diamond, the nitrogen vacancy has garnered particular interest due to its excellent room-temperature spin coherence, while the silicon vacancy defect has a strong and narrow zero-phonon-line (ZPL) emission and only a weak phonon-assisted sideband. Both defects' luminescence appears in the visible portion of the spectrum. For long-distance communications and applications with optical interfaces, photonic signals in the near-infrared range between 1.2 $\mu$m and 1.6 $\mu$m are most desirable, since scattering losses are greatly reduced and low absorption in silica optical fibers enables long-distance photon transmission. Furthermore, this wavelength lies in the range of the biological transparency window. 

Silicon carbide (SiC) has recently been discovered as a promising host for quantum bits and single-photon sources based on impurities. Its large bandgap ($>3\,$eV) gives ample room for strong transitions in the optical domain, while both silicon and carbon are predominantly free of nuclear spin by natural isotopic abundance. The crystal therefore offers a quiescent magnetic environment, and its hardness gives rise to high-frequency phonon modes which are only modestly occupied, even at room temperature. The recent study of optically active defects in silicon carbide has revealed several defects with properties of interest for quantum applications \cite{Weber2010}. Earlier work was focused on the electronic properties of impurities, and included detailed research on metallic dopants, of which several possess optical transitions in the near-infrared \cite{Baur1997}. Detailed studies of these defects have only recently resumed in light of the emergence of quantum-enhanced applications. For instance, the chromium lines at 1042 nm and 1070 nm were examined recently, showing promise as a spin-photon interface \cite{Koehl2017}. However, the long optical lifetime indicated a modest optical cross-section, presenting an obstacle for efficient photonic interactions.\\
Vanadium (V) is one of the most scrutinized defects in SiC, due to its amphoteric properties, which were routinely used for compensation of impure crystals. First optical studies showed strong and sharp luminescence lines, as well as narrow electron spin resonance (ESR) features with a linewidth smaller than 2 Gauss in an ensemble \cite{Schneider1990}. This width corresponds to a spin coherence lifetime of several tens of nanoseconds, enabling coherent spin manipulation. The defect was further found to have multiple charge states, which are of interest for spin-to-charge conversion and nuclear spin memory protection \cite{Elzerman2004,Saeedi2013}. \\
In this work, we have studied substitutional vanadium in silicon carbide with a view on applications in quantum information and communication. We investigate the optical properties of the neutral charge state, V$^{4+}$. This defect presents narrow optical emission lines in the telecommunications window near 1300 nm, an optical decay lifetime below 50 ns and a rich level structure. The defect therefore shows promise for photonic quantum technology.
We first describe its spectral features in sections \ref{subsec:spectrum} and \ref{subsec:lifetimes}. A theoretical picture of the defect properties is given in \ref{sec:model}.

\begin{figure*}
\centering
\includegraphics[width=2\columnwidth]{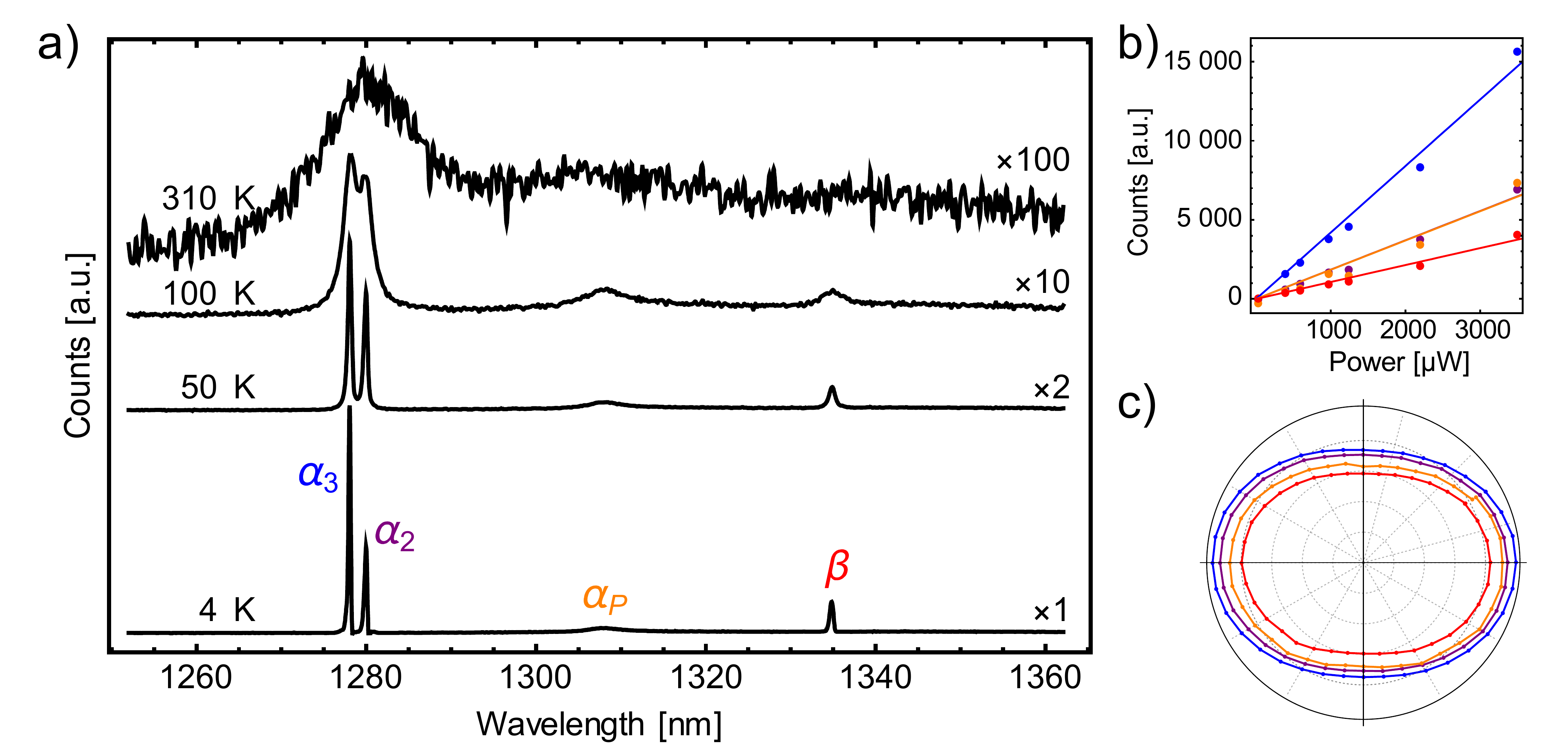}
\caption{a) Recorded spectrum of the luminescence at selected temperatures. b) Power dependence of the luminescence at 3.5 K, showing no saturation in the measured regime. c) Polarization dependence of the four different features, signals are re-scaled for clarity.}\label{fig:specPic}
\end{figure*}

\section{Experimental results}\label{sec:results}

\subsection{Setup and materials}\label{subsec:setup}

We examined a single-crystal 4H SiC sample with an homoepitaxial layer grown on an on-axis substrate \cite{Karhu2019}. The epilayer is 25 $\mu m$ thick and was grown with vanadium tetrachloride as a source of V-dopant, leading to an inclusion density of $2\times 10^{16}$ V/cm$^3$ as measured by SIMS. All photoluminescence (PL) measurements were performed using a 441 nm CW or pulsed laser source. The fluorescence was collected through the cryostat window with a 100x microscope objective (Mitutoyo M-Plan Apo, N.A.=0.7). Spectra were recorded using a grating spectrometer with a liquid-nitrogen-cooled detector, with a nominal resolution of 0.2 nm/pixel. All wavelengths are vacuum values. The fluorescence decay signals were recorded using a superconducting nanowire single photon detector (EOS 210 OS by Single Quantum). The fluorescence signals from the different spectral features were isolated using narrow bandpass filters with a full-width at half maximum smaller than 50 nm, rotated to select the desired wavelength range.
\subsection{Spectrum and polarization}\label{subsec:spectrum}
\begin{figure*}[htb!]
\centering
\includegraphics[width=2.\columnwidth]{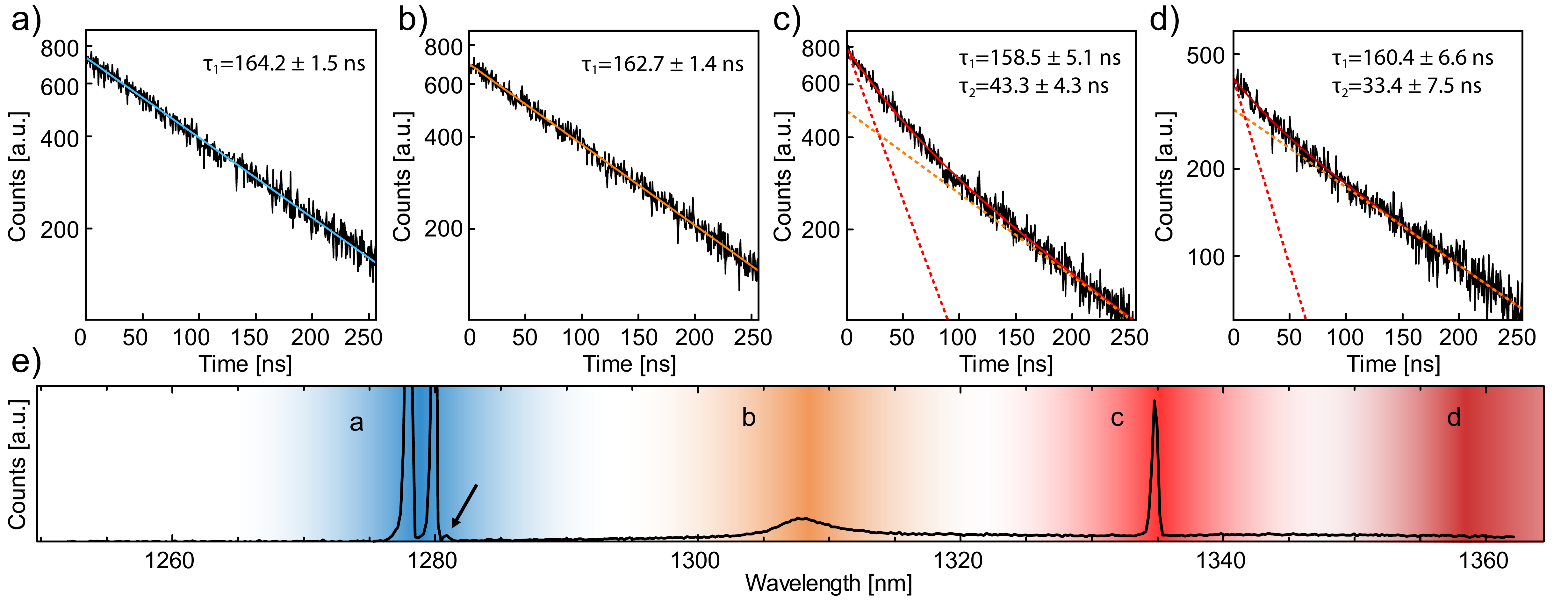}
\caption{Fluorescence decay for pulsed blue illumination measured at $\sim$4 K temperature. a)-d) Decay traces for filtered fluorescence at increasing wavelength ranges around 1280 nm, 1310 nm, 1334 nm and 1360 nm, respectively. The fitted decay functions are shown as solid lines. The dashed lines in c) and d) display the individual fitted decay components. 
e) Detail of the low-temperature spectrum. The four coloured bands indicate the filtering bands for a)-d). The arrow indicates the spectral feature used for estimation of the spin-orbit splitting in the excited state.}\label{fig:decayPic}
\end{figure*}
The spectrally resolved fluorescence [Fig \ref{fig:specPic} a)] under blue excitation of the defects reveals three sharp lines, a doublet around 1279 nm ($\alpha_3$ and $\alpha_2$) and a smaller feature at 1334 nm ($\beta$). The doublet has been attributed to the hexagonal site, the single line to the quasi-cubic site in 4H SiC. The broad feature between these lines near 1310 nm ($\alpha_P$) has not been assigned yet, but it has been hypothesized that it may correspond to a low-lying vibronic mode of the defect. The width of the $\alpha$ lines at temperatures below $30\,$K is below the spectrometer resolution ($\sim 35\,$GHz), while the $\beta$ feature is of approximately that width. Previous measurements in 6H-SiC however indicate that an unresolved doublet may give rise to the observed $\beta$ linewidth \cite{Kunzer1993}.\\
At low temperature, the $\alpha$ doublet makes up over $50\,\%$ of the recorded luminescence. It is however known that further phonon-assisted luminescence features appear beyond the spectral window recorded here, extending up to $\sim 1530\,$nm. We therefore estimate conservatively that the ZPL accounts for about one third of the defects' luminescence.\\
The doublet displays a temperature-dependent luminescence ratio. At 4 K, the $\alpha_3$ peak accounts for almost 70 \% of the doublet's luminosity, but the ratio approaches equilibrium approximately exponentially and is near unity at 100 K, above which temperature the lines are broadened beyond resolvability [Fig. \ref{fig:specPic} a)].\\
The dependence of the luminescence on excitation power is consistent with a linear behaviour within the available excitation range ($3.5\,$mW in front of the objective) [Fig \ref{fig:specPic} b)]. We therefore expect that high luminosity can be achieved with significantly greater excitation power, or for resonant driving. The origin of the slight ellipticity in the emission is as yet unknown. Importantly however, the polarization dependence of the luminescence displays near-identical behaviour across the spectrum  [Fig. \ref{fig:specPic} c)]. There are no indications for a strong spin dependence of the charge recapture, so it can be assumed that the excited-state spin sub-levels are populated equally and with random phases in the excitation process.\\
The strength of the overall luminescence shows no discernible dependence on temperature up to 40 K, then decreases slowly by $\sim$75 \% up to 310 K.\\

\subsection{Fluorescence decay lifetimes}\label{subsec:lifetimes}

\begin{figure}[htb!]
\centering
\includegraphics[width= \columnwidth]{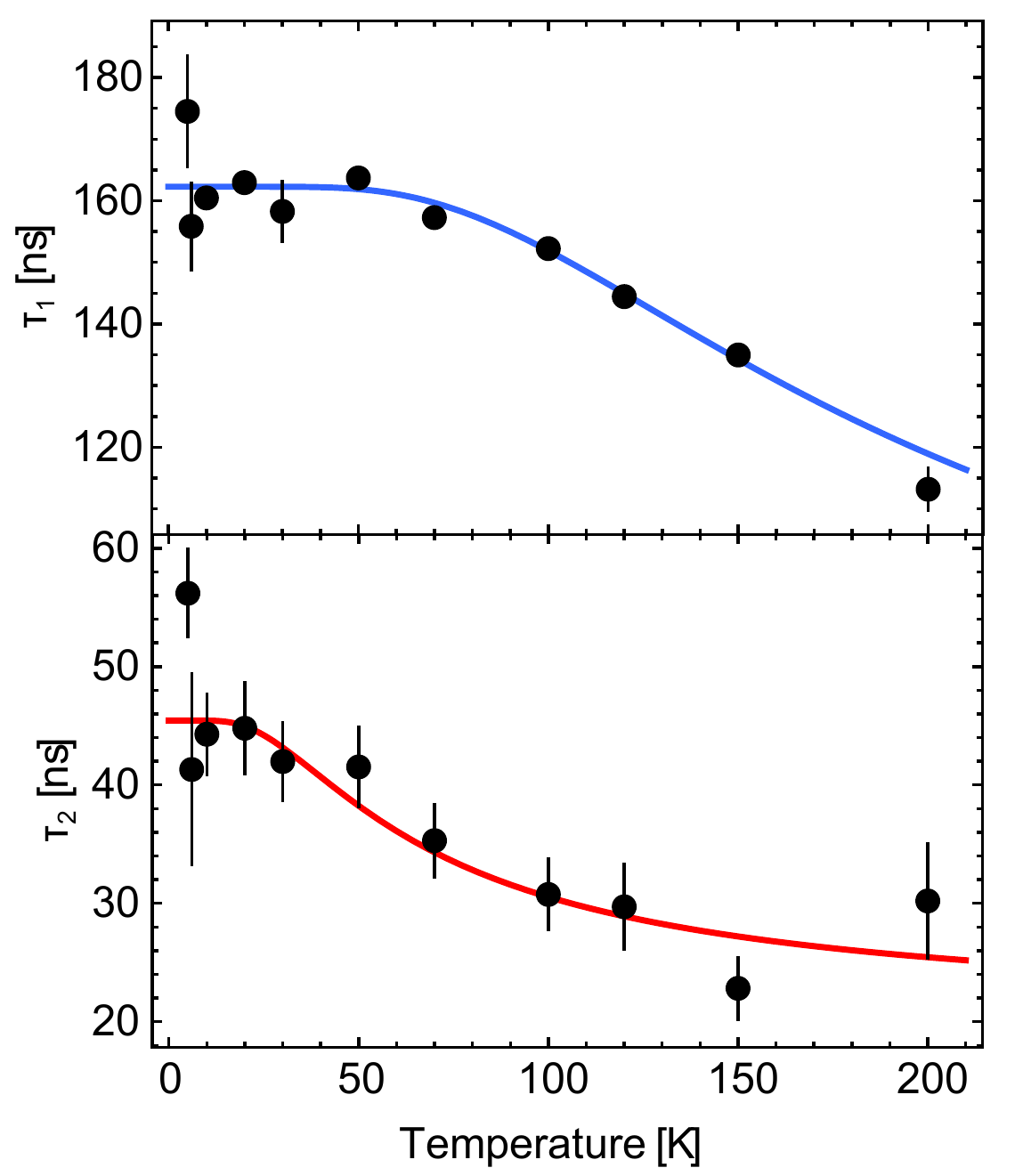}
\caption{Measured fluorescence decay lifetime versus temperature for the slower ($\alpha$) and faster ($\beta$) decay processes observed across the spectrum. The lines show fits to equation \ref{eq:decayVSTemp}.}\label{fig:decayVStempPic}\end{figure}

The decay traces for the two types of defect illuminated with a pulsed blue laser are shown in Fig. \ref{fig:decayPic}. A portion of the trace before the laser pulse is used to calculate the background counts. The data for the $\alpha_{2,3}$ doublet (black line) shows very good agreement with a single exponential decay (blue line) with a lifetime of $\tau_1=(164.2\pm 1.5)\,$ns, and a similar fluorescence lifetime is observed for the $\alpha_P$ feature. This observation confirms the assignment of the $\alpha_P$ feature to a low-lying phonon branch of the $\alpha$ defect. The  $\beta$ line requires a double exponential,
\begin{equation}
y(t)=A_1 \exp (-t/\tau_1) + A_2 \exp (-t/\tau_2),
\label{eq:doubExp}\end{equation}
to reach agreement with the data. While the magnitudes $A_1$ and $A_2$ of the decay components are similar, the  lifetimes differ significantly with $\tau_2=(158.5\pm 5.1)\,$ns and $\tau=(43.3\pm 4.3)\,$ns, where the error margins give the $3\sigma$ uncertainty of the measured value. A similar behaviour is observed in the spectral region  around $1360\,$nm. These characteristics indicate that local phonon modes give rise to a broad phonon sideband that extends from the $\alpha$ ZPL peaks to beyond the $\beta$ feature. The double-exponential behaviour around the $\beta$ emission peak is then due to the addition of $\alpha$-related, phonon-assisted luminescence and $\beta$ ZPL photons, while at longer wavelengths it is caused by the addition of the phonon sidebands of the two defect types.\\
The decay rates show no statistically significant dependence on temperature up to 20 K for the $\beta$ decay channels and up to 50 K for the $\alpha$ lines. At higher temperatures, the luminescence decay accelerates significantly, as shown in Fig. \ref{fig:decayVStempPic}.  The data points show weighted averages of decay times extracted from traces at different points in the spectrum. No spectral dependence of the decay lifetimes was discernible. We fit the observed decay rates to a model of thermally activated, non-radiative decay process, given by
\begin{equation}
\tau_{\text{tot}}(T)=\left[\frac{1}{\tau}+\frac{1}{\tau_{p}}\exp\left(-\frac{E_p}{k_B T}\right)\right]^{-1}.
\label{eq:decayVSTemp}\end{equation}

We find activation energies $E_p$ of $(28\pm 2)$ meV and $(8\pm 3)$ meV for the $\alpha$ and $\beta$ branches, with thermally-assisted decay lifetimes $\tau_p$ of $(83\pm 15)$ ns and $(36\pm 11)$ ns, respectively. These processes are likely related to thermally induced charge transfer processes involving other impurities. Their relatively high activation energies nonetheless underline that both defects can be expected to display stable features at temperatures below 10 K.

%\textbf{Red illumination}
%The fluorescence decay was also recorded using a red laser, albeit with significantly lower power (see Fig. \ref{fig:decayPicRed}). The low count rate prohibited the collection of statistically significant traces. The extracted decay lifetimes are too uncertain to warrant comparison with the previously recorded traces. Nonetheless, the available data indicate no disagreement with the data from blue excitation, namely a single exponential decay for the $\alpha$-site and a double exponential for the $\beta$-site. 
\section{Theory and numerical results}\label{sec:model}

\subsection{Defect model and details of calculations}

Previous results showed \cite{IvadyPRL2011} that vanadium substitutes Si site in 4H SiC. Since two inequivalent substitutional sites, $k$ and $h$, exist due to the particular packing of Si-C bilayers along the c-axis in 4H SiC, vanadium will form two distinct color centers in 4H SiC. We label the two distinct configurations as vanadium ($k$) and vanadium ($h$) accordingly. We study the electronic structure and optical properties of these configurations by \emph{ab initio} calculations for identification of the vanadium-related $\alpha$ and $\beta$ emitters.

Calculations were carried out in the framework of Kohn-Sham (KS) Density Functional Theory (DFT) by means of the Perdew-Burke-Ernzerhof (PBE) \cite{PBE} functional. We rationalize the choice of this functional in Sec.~\ref{ssec:subsDFT}. 
The vanadium ($k$) and ($h$) defects were modeled in a 576-atom supercell. The Kohn-Sham wavefunctions were expressed by plane waves with a cutoff of 420~eV where the ions were treated by projector-augmented wave (PAW) \cite{PAW} potentials as implemented in \textsc{vasp} code \cite{VASP}. The large supercell suffices to apply $\Gamma$-point for sampling the Brillouin-zone which also allowed us to carefully follow the degeneracy of the defect levels in the fundamental band gap. The geometries were relaxed by minimizing the total energy with respect to the coordinates of the atoms until the forces fell below using the threshold of 0.01~eV/\AA. 
%In the case of charged defects Freysoldt correction \cite{Freysoldt} was applied in the total energy .

The excited state is calculated by $\Delta$SCF method~\cite{Gali2009}, in order to determine the zero-phonon-line (ZPL) energy. In order to calculate the phonon sideband, the Huang-Rhys (HR) approximation was employed as implemented by Gali \textit{et al.} \cite{galiNatComm}. The strength of the coupling is represented by the total HR factor ($S(\hbar\omega$)). From $S(\hbar\omega$) spectral function ($A(E_{\text{ZPL}} - \hbar\omega)$) of the photolominescence spectrum can be calculated. Detailed derivation of $S(\hbar\omega$) and $A(E_{\text{ZPL}} - \hbar\omega$) can be found in \cite{galiNatComm} and \cite{alkauskasIOP}. 
The contribution of the ZPL emission to the total emission containing the phonon sideband is the Debye-Waller factor (DW) that can be readout in the experimental PL spectrum which is related by DW$= \text{exp}\{S(\hbar\omega)\}$ to the calculated HR factor, $S$.

\subsection{Test on the applied semilocal DFT functional}
\label{ssec:subsDFT}

The HSE06 range-separated functional developed by Heyd, Scuseria and Ernzerhof \cite{HSE06} was proven to be a powerful tool to study point defects in solids \cite{Deak2010, Freysoldt2014}. However, it has been shown that transition metal defects can introduce strongly localized orbitals (e.g., $d$ electrons) which is significantly different from the character of the $sp^3$ electron bath of the host semiconductors, and HSE06 fails to simultaneously describe both states. To remedy this issue, Iv\'ady \textit{et al.} \cite{ivadyDFT1} developed a technique in which the HSE06 functional is corrected by an orbital dependent term inspired from DFT+$U$ method, that is called HSE06 + $V_w$ method. The additional term reads as
\begin{equation}
V_w = w\bigg(\frac{1}{2} - n\bigg),
\label{formeq}
\end{equation}
where $w$ represents the strength of the potential and $n$ is the occupation number. The value of $w$ can be determined via the generalized Koopmans' theorem (gKT)\cite{lanygKT}. We carried out calculations in order to determine the potential strength, which yielded $w = 2.2$~eV for both vanadium ($k$) and ($h$) defects, indicating identification of the accurate functional for the vanadium impurity in 4H SiC. We use this functional to test the computationally less costly semilocal PBE functional.
\begin{figure*}[t!]
\centering
\includegraphics[width=2\columnwidth]{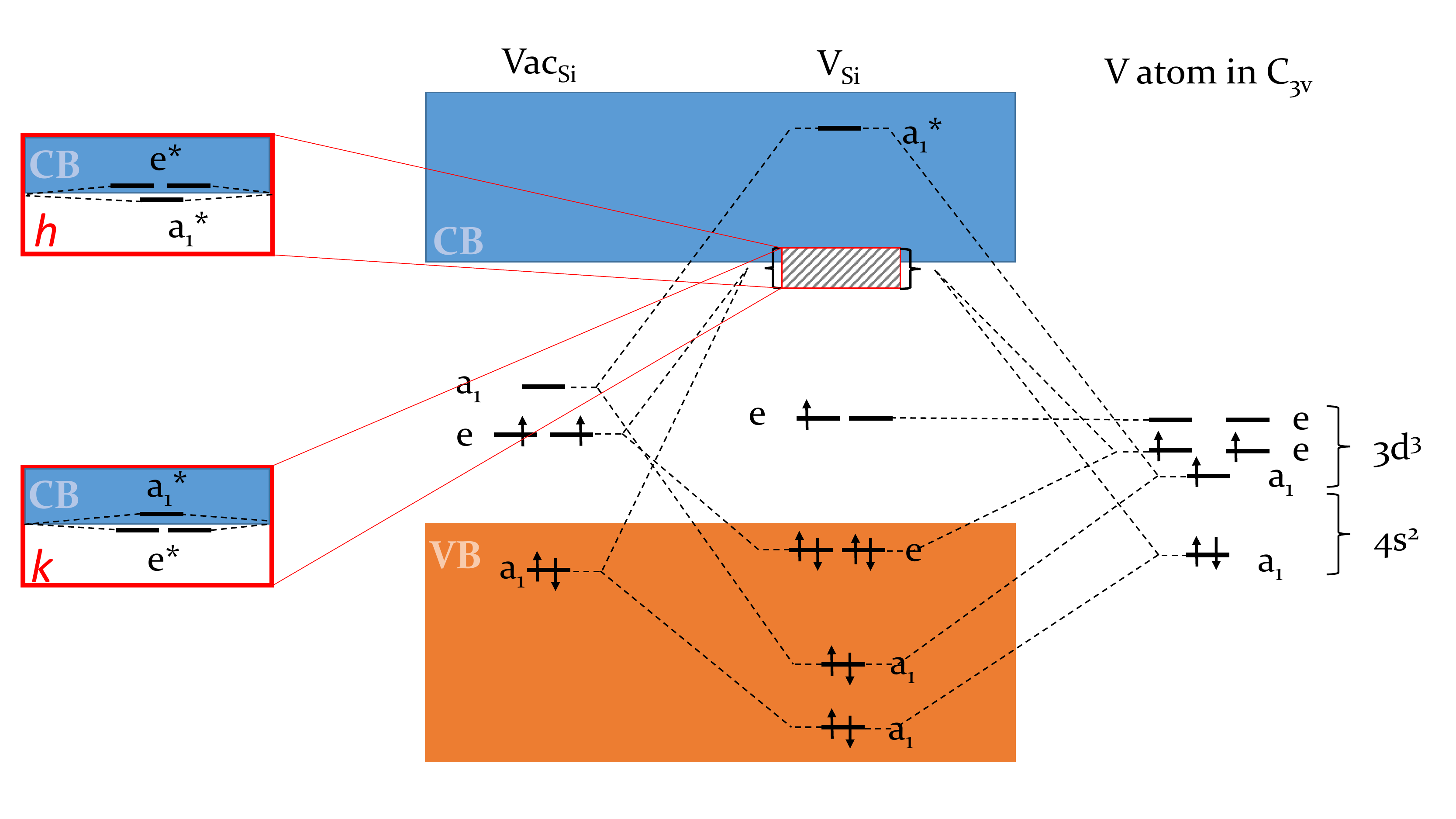}
\caption{Defect molecular orbital diagram for vanadium ($h$) and vanadium ($k$) in 4H SiC under C$_\text{3v}$ symmetry regarding only the crystal-field splitting. Characters of all Kohn-Sham levels are indicated. Antibonding levels are denoted by an asterisk. Vac$_{\text Si}$ indicates the silicon vacancy levels.  CB (VB) indicates the conduction (valence) band.}
\label{fig:KSLevels}
\end{figure*}
We found that the order of defect levels for the two defect configurations agrees by the two methods (cf. Fig. \ref{fig:KSLevels}). Furthermore, the localization of defect wavefunctions on vanadium atom agrees within $\sim$10\% as obtained by the two methods. We concluded that PBE functional can provide semi-quantitative results for vanadium in 4H SiC, thus we employed this DFT functional in the context. However, this method is not expected to provide the correct absolute value for the energy of the excited state of the defects, for which we instead rely on the experimental data.
\subsection{Identification of emitters and calculated photoluminescence spectrum}
Vanadium is neutral in a wide range of doping levels \cite{ivadyDFT2}, thus it is assumed that the $\alpha$ and $\beta$ emission lines originate from the neutral vanadium defects \cite{Mitchel2007}. The observed spectral features support this model \cite{Baur1997}, which assumes a spin doublet ground state. We considered the neutral vanadium defect accordingly. 
\begin{figure*}[t!]
\centering
\includegraphics[width=2\columnwidth]{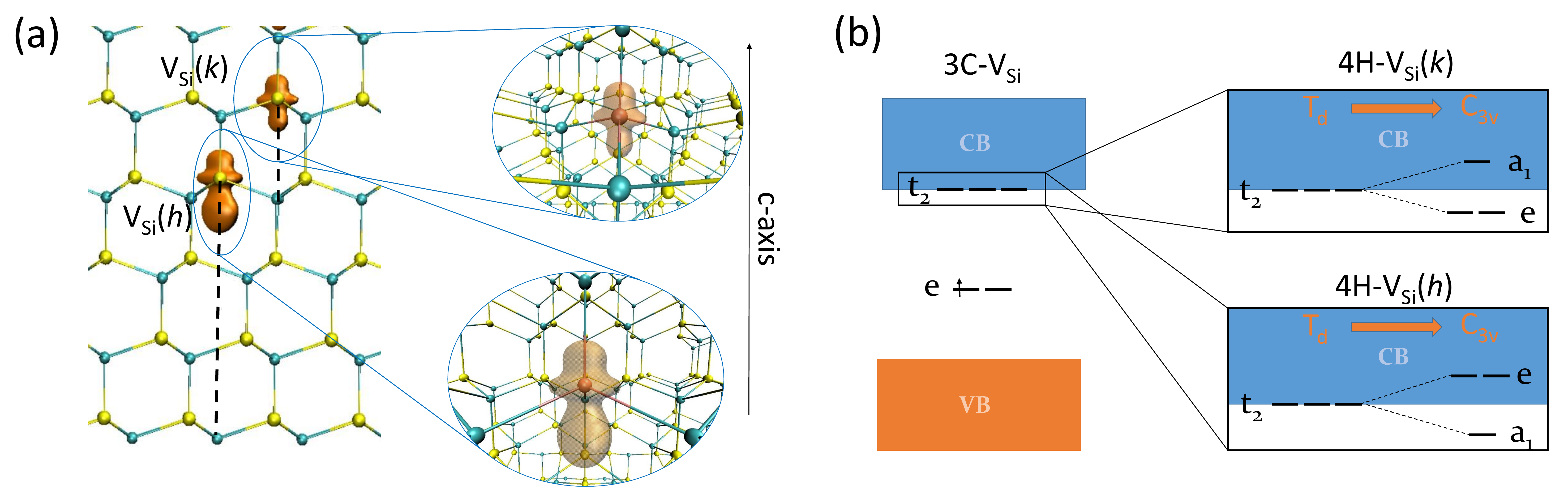}
\caption{ (a) Defect models of vanadium ($h$) and ($k$) depicted in the same supercell in orthographic view. Isosurfaces --- using the same isovalue --- of spin densities are illustrated for both defects. Dashed lines illustrate the distances of the nearest neighbor atoms in the $z$-direction as the origin of the potential confinement along the $c$-axis. (b) Splitting of $t_2$ level under C$_\text{3v}$ symmetry. Both possible energy orderings of the splitting are depicted with the assignment of the corresponding vanadium defects in 4H SiC. VB and CB are the valence and conduction bands, respectively.}
\label{fig:a1}
\end{figure*}
Application of group theory can be very useful in determining the electronic structure of the defect. In Fig. \ref{fig:KSLevels} the defect-molecular orbital diagram of the vanadium defect is depicted which assumes an interaction between the orbitals of Si-vacancy and vanadium atomic orbitals. According to group theory considerations the electronic structure introduced of the neutral vanadium defect is $a_1(2)a_1(2)e(4)e(1)$ whereas the empty states are $a_1^*$ and two $e^*$. The calculations imply that all the fully occupied states fall in the valence band (VB) and only the singly occupied atomic-like $e$ levels appear in the gap that constitutes $^2E$ ground state. One of the empty $e^*$ lies high in the conduction band (CB) whereas the $a_1^*$ and the other $e^*$ are resonant with the conduction band edge. Our DFT calculations revealed that the order of these $a_1^*$ and $e^*$ levels depends on the defect: 
$e^*a_1^*$ and $a_1^*e^*$ for the $k$ and $h$ sites, respectively (see Fig. \ref{fig:KSLevels}).

To understand this effect, it is illustrative to derive the electronic structure from that of vanadium defect in 3C SiC exhibiting T$_{\text{d}}$ symmetry. Accordingly, vanadium atom introduces a double degenerate $e$ level in the band gap and a triple degenerate $t_2^*$ level resonant with the CB edge. Under C$_\text{3v}$ crystal field of the defect in 4H SiC the $t_2$ state splits into a double degenerate $e$ and non-degenerate $a_1$ level. The energy order of these states can only be determined by a closer inspection of the defect environment [cf. Fig. \ref{fig:a1} (b)]. According to the calculations, the $e^*$ state exhibits negligible extension in along $c$-axis whereas $a_1^*$ state ($d_{zz}$-like orbital) is sensitive to the environment along the $c$-axis. Because of the crystal structure of 4H SiC, the $a_1$ state is more confined along the $z$-direction in ($k$) than that in ($h$) configuration [dashed lines in Fig. \ref{fig:a1}(a)]. The stronger confinement at ($k$) site over ($h$) site pushes up the empty $a_1^*$ level at vanadium ($k$) with respect to that at vanadium ($h$) whereas the corresponding $e^*$ levels are the same.

These differences have considerable consequences for the nature of the PL emission from these defects. The emission comes from the $^2E$ excited state at ($k$) site whereas it comes from $^2A_1$ at ($h$) site. This influences the fine structure of the excited state and the polarization of the emitted photons. Furthermore, the zero-phonon-line (ZPL) energy should be larger for ($k$) configuration than that for ($h$) configuration if we assume the same Stokes shift in the emission process. 

PBE DFT calculations indeed result in 0.845~eV ZPL energy at $(k)$ site and 0.785~eV ZPL energy at ($h$) site. Here, we neglected the spin-orbit splitting and used static Jahn-Teller distorted structures in the total energy calculations. The absolute values in the calculated ZPL energies are within 0.1~eV. More importantly though, the calculated ZPL energy difference with this simple method is 0.06~eV which is close to the experimental value of 0.04~eV. These results strongly imply that $\alpha$ and $\beta$ emitters are the vanadium defects at $(k)$ and ($h$) sites, respectively, in contrast to the present assumptions in the literature~\cite{Schneider1990}.

\begin{figure}[htb!]
\centering
\includegraphics[width=1\columnwidth]{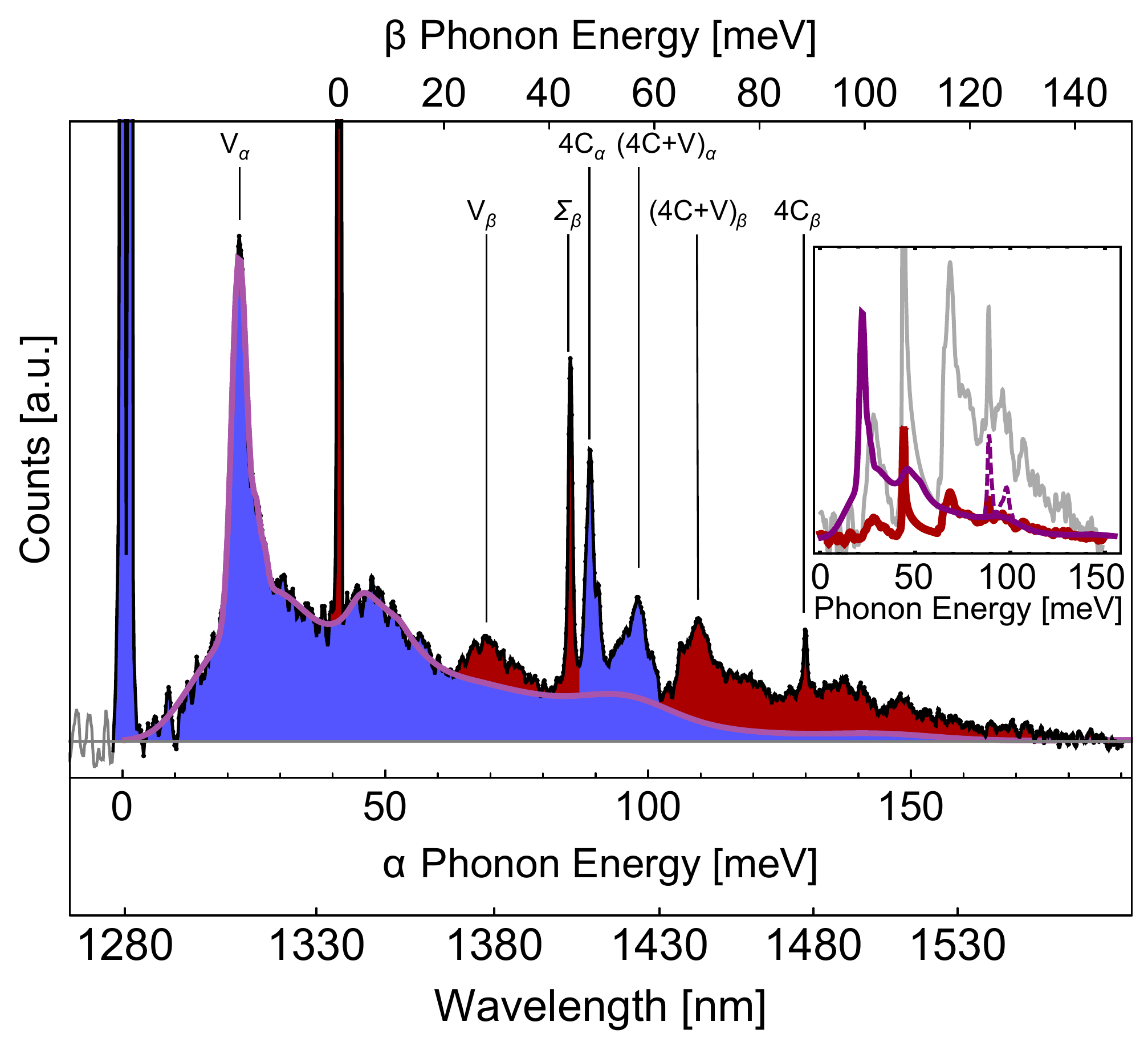}
\caption{Detail of the phonon sideband, with assignments extracted from the \emph{ab initio} defect models. The purple line shows a fit to the features with energy below 64 meV which are assumed to originate solely from the $\alpha$ defect. The blue (red) shaded area is assigned to the $\alpha$ ($\beta$) line. Inset: fitted $\alpha$ phonon energy (purple line) compared to the extrapolated and shifted $\beta$ features (red line, unscaled; gray: scaled to equal ZPL intensity). The lines are smoothed and interpolated for clarity. The dashed segment is the overlayed 4C$_\alpha$ and (4C+V)$_\alpha$ portion of the recorded PSB.}\label{fig:phonons}
\end{figure}

Examination of the phonon sideband (PSB) of the vanadium emitters is of great importance as on the one hand, it gives further insight into the optical properties of the defects, and on the other hand, it presents pathways for off-resonant excitation. The average Debye-Waller (DW) factor of the two defects can be evaluated directly from the measurement and yields $\overline{\text{DW}}=34\,$\%. We can further calculate limits on the DW factors of the two defects by partitioning the spectrum at the $\beta$ PSB, which yields $31~\% <$DW$_\alpha <60\,$\%, and DW$_\beta >10\,$\%. To gain further understanding of the phononic properties of the two emitters, we construct a numerically optimized spectrum to match the observed photoluminescence (see Fig. \ref{fig:phonons}). Guided by the \emph{ab initio} model, we assume that the $\beta$ defect has no significant PSB emission with phonon energies smaller than $\sim 20\,$meV. We furthermore impose that the PSB emission be zero for phonon energies greater than 200 meV. We then construct a summed spectrum out of five Gaussian sidebands of the form
\begin{equation}
I(\delta)=I_0\sum_{j=1}^{j=10}\frac{1}{\sqrt{j\pi}\sigma}\exp\left[-\left(\frac{\delta-\Delta_0}{\sqrt{j}\sigma}\right)^2\right].
\label{eq:phonons}
\end{equation}
Each sideband is then turned into a weighted doublet to account for the $\alpha_2$ and $\alpha_3$ emission, by appropriate adjustment of the factors $\Delta_0$ and $I_0$ to match the observed 1.47 meV splitting and strengths in the $\alpha$ ZPL. The remaining PSB luminescence is mostly assigned to the $\beta$ defect, aside from a prominent set of features near the $\alpha$ phonon energy band. The calculated spectrum allows us to refine our estimates to DW$(\alpha)~39\,$\% and DW$(\beta)\sim 22\,$\%.
  
The table below lists the calculated Huang-Rhys $S$ and the derived Debye-Waller factors, as well as the calculated radiative lifetime $\tau_{\text{rad}}$. Comparing this value to the measured photoluminescence lifetime ($\tau$ (exp.)) yields an estimate of the radiative efficiency $\eta_{\text{rad}}$. A lower limit for the total efficiency of ZPL transition, $\eta_{\text{tot}}$, is then calculated as the product of $\eta_{\text{rad}}$ with the estimated Debye-Waller factors. 
\begingroup 
\squeezetable
\begin{table} [h!]
\caption{\label{tab:radProperties} Summary of measured and calculated radiative properties for the ($h$)- and ($k$)-configurations of V$^{4+}$ in 4H SiC. (th.) and (exp.) indicate calculated and measured values, respectively. $\tau_{\text{rad}}$, $\tau_{\text{tot}}$ (exp.) and $\tau_{\text{NR}}$ are given in ns. $\tau_{NR}$, $\eta_{\text{rad}}$, $\eta_{\text{tot}}$ are extracted from the preceding measured and calculated values.}
\begin{ruledtabular}
\begin{tabular}{c c c c c c c c c c} 
site & S (th.)& DW (th.) & DW & $\tau_{\text{rad}}$ (th.)& $\tau_{\text{tot}}$ (exp.) & $\tau_{\text{NR}}$ & $\eta_{\text{rad}}$ & $\eta_{\text{tot}}$ \\  
k & 0.66 & 0.52 & 0.39 & 704 & $\tau_1=$163 & 212 & 23 \% & 8.9 \% \\ 
h & 0.79 & 0.45 & 0.22 & 277 & $\tau_2=$43 & 47 & 15 \% & 3.3 \% \\ 
\end{tabular} 
\end{ruledtabular}
\end{table}
\endgroup

The difference in the experimentally deduced and theoretical DW factors may arise from the sharp phonon peaks in the observed photoluminescence spectra, which are not reproduced by our simulations. These cannot be accounted for by Huang-Rhys theory, and may be due to an Herzberg-Teller mechanism \cite{Londero2018}. In particular, the very sharp $\beta$ feature marked $\Sigma_\beta$ in Fig. \ref{fig:phonons} cannot be assigned to a particular oscillatory mode. This feature however offers a strong absorption cross-section far off resonance ($\lambda=1274\,$nm) from the $\beta$ ZPL (1334 nm), which will be of benefit for photon source applications. Similarly, the 4C$_\alpha$ feature is expected to have a corresponding absorption feature near 1170 nm, and may even enable state-selective driving of the defect since the $\alpha_2$ and $\alpha_3$ components are at least partially resolved. 
As the energy gap is smaller in the ($h$)-configuration than in the ($k$)-configuration, the non-radiative lifetime of the ($h$)-configuration can be expected to be shorter. Its radiative lifetime is also shorter than that of the ($k$)-configuration. It is conceivable that the more de-localized excited wavefunction is responsible for this feature. We furthermore note that the use of blue excitation may lead to non-radiative recombination pathways that are excluded when using resonant excitation. It is therefore possible that a photo-luminescence excitation experiment may reveal longer excited-state lifetimes than those observed here, yielding more favorable transition efficiencies.

\subsection{Quantum optical properties}
\begin{figure}[htb!]
\centering
\includegraphics[width=1\columnwidth]{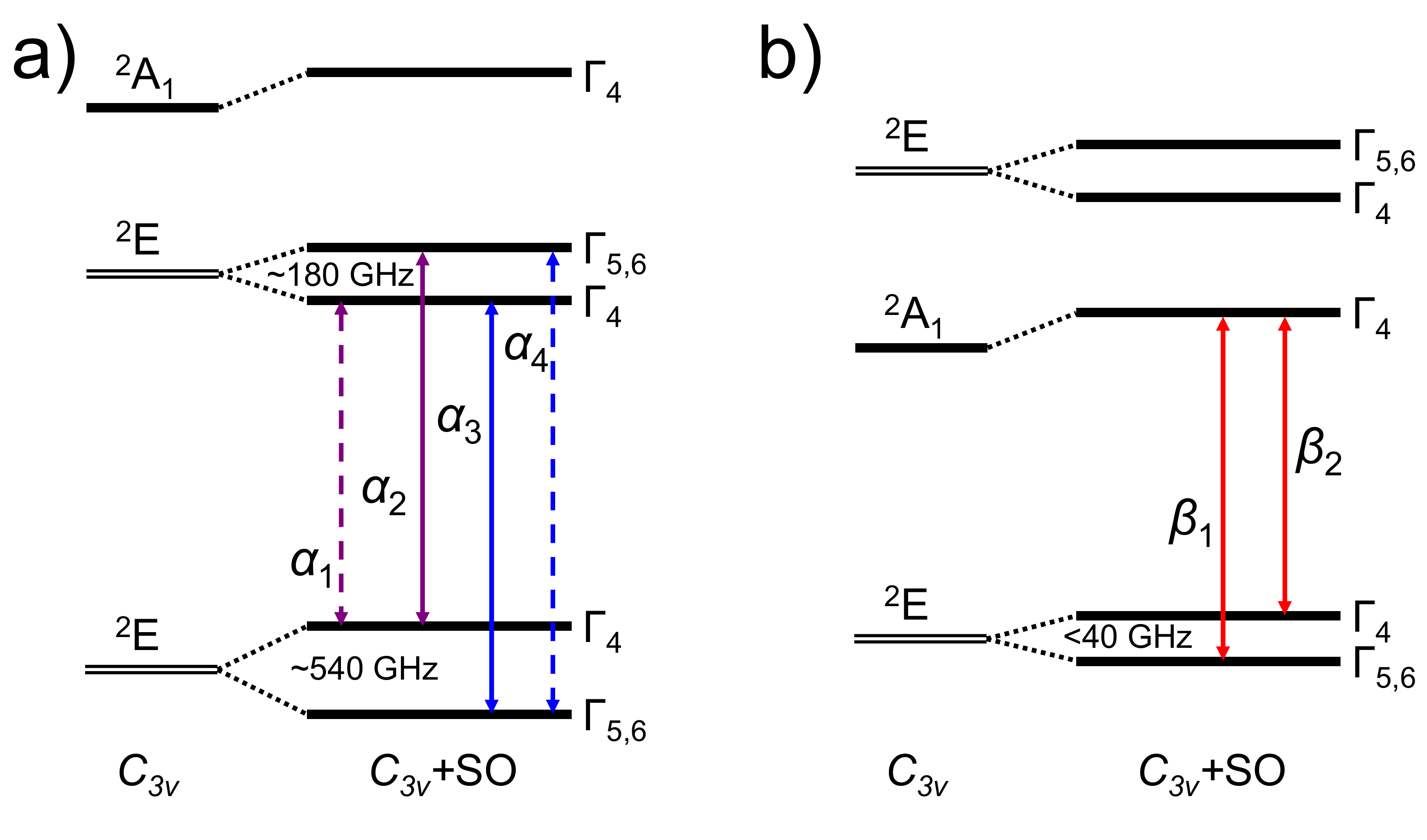}
\caption{Level structure with tentative assignment from group theory, simulation, previous work \cite{Kaufmann97} and measurement.  a) $\alpha$ lines originating from the cubic ($k$) site. The weak luminescence feature at slightly lower energy than the $\alpha_{2,3}$ doublet allows to estimate the spin-orbit (SO) splittings as shown (see Fig. \ref{fig:decayPic}). b) $\beta$ lines originating from hexagonal ($h$) site.}\label{fig:levels}
\end{figure}

While V in SiC shares some features with the silicon divacancy in diamond \cite{Rogers2014,Becker2016}, there are important differences beyond the beneficial optical frequency. As is shown in Fig. \ref{fig:levels}a\&b the spin-orbit (SO) splitting for the hexagonal $(h)$ site is comparable to the silicon divacancy SO splitting in diamond while the cubic $(k)$ site shows a more than ten times larger ground state SO splitting \cite{Kaufmann97}. The ground state is seated deeply within the band gap, enabling strong localization of the electron wavefunction. This points to long spin coherence lifetimes, which have indeed been observed in ensembles \cite{Schneider1990}. Furthermore, together with the relatively large nuclear magnetic moment of $^{51}V$ ($S=7/2$, $g_N=+1.471$ $\mu_N$), the strong localization of the electron wavefunction enables a significantly larger hyperfine coupling frequency ($A_\|=235$ MHz) than the ESR linewidth \cite{Baur1997,Stone2014}. The hyperfine octet furthermore offers a large Hilbert space for quantum information storage and manipulation.\\
An important application for these defects will be the generation of single photons, which can be enhanced in an optical resonator. From the measurements and our estimate $\eta_{\text{tot}}\sim 8.9\,\%$ of the transition efficiency for the $\alpha$ doublet, we can estimate the cooperativity that can be achieved in a Fabry-P{\'e}rot resonator:
\begin{equation}
C=\frac{2}{\pi}\frac{\sigma_E}{\sigma_C}\eta_{\text{tot}}\frac{f_L}{n_{\text{SiC}}}F.
\label{eq:cooperativity}\end{equation}
Here $\sigma_E=3\lambda^2/2\pi$ is the ideal optical cross-section of the emitter, while $\sigma_C=\pi w_C^2$ is the cross-section of the cavity mode with the $1/e^2$ mode field radius $w_C$. The factor $f_L=(L_{vac}+n_{\text{SiC}}L_{\text{SiC}})/(L_{\text{vac}}+n^2_{\text{SiC}}L_{\text{SiC}})$ accounts for the modification of the electric field strength in a heterogeneous cavity partially filled with SiC, and assumes anti-reflection coating on the inner surface. We assume $L_{\text{vac}}=L_{\text{SiC}}=5\,\mu$m. The enhanced photon emission into the cavity mode has a probability that depends on the cooperativity as with $\eta_{\text{cav}}=2C/(2C+1)$. With current, published values for microcavities in the telecom regime ($1.3\,$mm radius of curvature, $F(\text{max})=3.4\times 10^4$ \cite{Kuhn2017}), this estimate yields a value of $\eta_{\text{cav}}=82\,\%$, while optimizing the mirror reflectivity for photon extraction gives an output efficiency of greater than $50\,\%$. Reduction of the cavity dimensions and an increase in finesse could realistically lead to output efficiency values exceeding $90\,\%$. These values indicate that V in SiC may be used to produce fast and highly efficient, cavity-enhanced single-photon sources, and that quantum state readout \cite{Hanks2017} can be drastically improved in such a setting. Optical microresonators integrated directly in SiC offer an alternative path for optical enhancement \cite{Calusine2016,Bracher2017}. 
\section{Discussion}\label{sec:discussion}
The optical properties of the studied vanadium defects indicate that they are strong candidates for photonic applications. The recorded spectrum is in agreement with the literature. However, our \emph{ab-initio} simulations suggest that the site assignment of the two defects should be reversed: The $\alpha$-lines originate from the cubic site, while the $\beta$ lines are emitted from the hexagonal site. The fluorescence decay is remarkably fast for an intra-shell $3d$ transition.

The double-exponential decay recorded in the spectral region of the $\beta$ line most likely results from the sum of phonon-assisted $\alpha$ luminescence and a faster optical transition in the $\beta$ defect. Taken together, the shorter lifetime and the lower fluorescence count rate could be caused by stronger non-radiative or phonon-assisted decay processes for the $\beta$ defect, but the lower count rate could also be due to a reduced recombination efficiency in the excited state, since the electron wavefunction is more strongly localized at this site. \\
Future efforts the theoretical focus should be placed on deepening our understanding of the spin-orbit structure and of the possible Herzberg-Teller features in the PSB. Resonant spectroscopy will be required to shed further light on the nature of the decay processes, including non-radiative decay channels and thermally-enhanced decay rates.\\ 
Both sites present a rich level structure that will enable coherent optical manipulation and spin-selective optical transitions, in turn leading to applications in quantum communication and computation \cite{Fleischhauer2005,Nemoto2014,Zwier2015,Yale2016}. These mechanisms can be further enhanced by placing the defects into optical microcavities, where the long wavelength will be beneficial in reducing losses due to scattering. Furthermore, the transitions both lie in the transparency window of silicon, opening the possibility of integration with photonic, mechanical and electronic structures in the most developed semiconductor platform \cite{Derntl2014,Rickman2014}. Furthermore, the emission lies in the second biological window, and may therefore be of interest for \emph{in-vivo} bioimaging.
In summary, we have studied the luminescence of neutral substitutional vanadium centres in 4H SiC. Temperature- and time-dependent measurements have allowed us to shed light on previously unassigned features of the spectrum, and have revealed attractive features including a small inhomogeneous linewidth and fast optical transitions. The characteristics of the vanadium centres studied herein are reminiscent of the molybdenum defect in SiC and the silicon-vacancy complex in diamond\cite{Bosma2018,Rogers2014,Becker2016}, but at technologically practical wavelengths in the telecommunication "O"-band, and favourable spin properties. The centres therefore merit further examination for applications in quantum technology. 

\textbf{Acknowledgements} Financial support was provided by the FWF project I 3167-N27 SiC-EiC, the OeAD project HU 12/2016 PERFEQT and the Carl-Trygger Stiftelse f\"or Vetenskaplig Forskning (CTS 15:339), the Swedish Research Council (VR 2016-04068), the Swedish Energy Agency (43611-1). The support from {\'U}NKP-18-3-I New National Excellence Program of the Ministry of Human Capacities of Hungary is acknowledged by A. Cs\'or\'e. A. Gali acknowledges the support from the National Research, Development and Innovation Office in Hungary (NKFIH) Grant Nos.\ 2017-1.2.1-NKP-2017-00001 (National Quantum Technology Program) and NVKP\_16-1-2016-0043 (NVKP Program).
\\
\textbf{Author contributions} L. S., T. F. and M. T. performed the measurements. A. C., G. T. and A. G. simulated the defect structure and performed group theory calculations. R. K. and J. U. H. produced the sample. M. T. initiated and coordinated the work, and wrote the manuscript. All authors contributed to data analysis and to the manuscript text.\\
\bibliographystyle{apsrev4-1}
\bibliography{references}{}

\end{document}